\documentclass[prb,twocolumn,showpacs,preprintnumbers,amsmath,amssymb]{revtex4}

\usepackage{graphicx}
\usepackage{dcolumn}
\usepackage{bm}

\begin{document}

\preprint{APS/123-QED}

\title{Self--consistent simulation of quantum wires defined by local oxidation of Ga[Al]As heterostructures}

\author{Christian P. May}
 \email{cmay@itp.phys.ethz.ch}
\author{Matthias Troyer}%
 \email{troyer@phys.ethz.ch}
\affiliation{%
 Institute for Theoretical Physics, ETH Zurich, CH-8093 Zurich, Switzerland 
}%

\author{Klaus Ensslin}
 \email{ensslin@phys.ethz.ch}
\affiliation{
 Solid State Physics Laboratory, ETH Zurich, CH-8093 Zurich, Switzerland
}%

\date{\today}

\begin{abstract}
We calculate the electronic width of quantum wires as a function of their lithographic width in analogy to experiments performed on nanostructures defined by local oxidation of Ga[Al]As heterostructures. Two--dimensional simulations of two parallel oxide lines on top of a Ga[Al]As heterostructure defining a quantum wire are carried out in the framework of Density Functional Theory in the Local Density Approximation and are found to be in agreement with measurements. Quantitative assessment of the influence of various experimental uncertainties is given. The most influential parameter turns out to be the oxide line depth, followed by its exact shape and the effect of background doping (in decreasing order).
\end{abstract}

\maketitle

\section{\label{intro}Introduction}
There have been several semiconductor quantum dot calculations assuming simplifications such as parabolic confining potentials\cite{Metzner, Lehmann, Pekka} and a number of self--consistent solutions of coupled Schr\"odinger and Poisson's equations in III/V semiconductor nanostructures.\cite{Stern, Snider1, Snider2, Ram-Mohan, Zhang, PRB.49.7474, PRB.47.13884, PRB.60.4485, JAP.95.5545, JAP.60.2680} 

However in particular the depletion mechanism in two--dimensional electron gases (2DEG) caused by local oxidation of the surface of AlGaAs/GaAs heterostructures with an atomic force microscope (AFM) has so far only been studied in simplified models assuming a non--realistic geometry.\cite{May1} By AFM induced oxidation,\cite{Held1,Fuhrer} confining walls can be defined with high accuracy enabling long quantum wires with only a few modes. Previous studies on quantum wires have either used analytical approaches,\cite{PhysRevB.33.8874} lacked self--consistency\cite{Sanchez} or applied simple geometrical assumptions.\cite{PhysRevB.47.4413} 

In this work, we perform a self--consistent numerical simulation of the effect of oxide lines on top of an AlGaAs/GaAs heterostructure assuming a realistic oxide line profile. In particular, a structure with two oxide lines defining a quantum wire is studied within the framework of Density Functional Theory (DFT)\cite{Hohenberg} assuming the Local Density Approximation (LDA)\cite{Kohn} and the results thereof are compared to available experimental data. The main point of this paper is to treat this specific kind of semiconductor nanostructure on a quantitative level, which has not been done before. Moreover, various uncertainties like background charges are assessed quantitatively in order to rank them by importance and judge their respective influences.
Obtaining results for this type of quantum wires which agree with experiments represents a major step towards a full self--consistent simulation of AFM lithography defined III/V semiconductor nanostructure devices such as quantum dots and quantum point contacts with realistic potentials.

\section{Methodology}
In order to solve the electrostatic problem, it is necessary to solve Poisson's equation. In our case of a two--dimensional calculation, it reads
\begin{equation}
  \nabla \left(\epsilon(x,y) \nabla \right) \Phi(x,y)= -\frac{\rho(x,y)}{\epsilon_0},
  \label{PoissonEq}
\end{equation} 
where $\epsilon$ represents the space--dependent dielectricity tensor and $\rho$ denotes the charge density, given by
the ionized donor/acceptor concentrations $N_n$, $N_p$ as well as the electron density $n$:
\begin{equation}
  \rho(x,y)=q(N_n(x,y)-N_p(x,y)-n(x,y)).
\end{equation}
Furthermore, we solve a one--particle Schr\"odinger equation
\begin{equation}
  \left[-\frac{\hbar^2}{2} \nabla \left(\frac{1}{m^{*}(x,y)} \nabla \right) + V(x,y) \right] \Psi(x,y) = E\, \Psi(x,y).
  \label{SchroedingerEq}
\end{equation}
In Eq. (\ref{SchroedingerEq}), $m^{*}$ stands for the space--dependent effective mass, while the potential $V$ consists of the band edge offsets $\Delta E_c$ at heterostructure interfaces, the electrostatic potential $\Phi$ and exchange and correlation terms, which are taken into account through the explicit parameterization $V_{\rm xc}$ given by Hedin and Lundqvist.\cite{Hedin}
\begin{equation}
  V(x,y)=-q\Phi(x,y) + \Delta E_c(x,y) + V_{\rm xc}(\rho(x,y)).
\end{equation}
The exchange--correlation potential explicitly reads
\begin{equation}
 V_{\rm xc}(\rho)=\frac{-2 \mathrm{Ry}^* }{\pi^{\frac{2}{3}} (\frac{4}{9})^{\frac{1}{3}} r_s}  \left( 1+0.7734 \frac{r_s}{21} \log{\left(1+\frac{21}{r_s} \right)} \right),
\end{equation}
where
\begin{equation}
 r_s(x,y)=\left( \frac{4}{3} \pi {a^*}^3 \rho(x,y) \right)^{-\frac{1}{3}}
\end{equation}
while $a^*$ and $\mathrm{Ry}^*$ are the effective Bohr radius and Rydberg constant in the respective material, which are given by
\begin{eqnarray}
  a^*(x,y) &=& 4 \pi \epsilon_0 \epsilon(x,y) \frac{\hbar^2}{m^*(x,y) q^2}\\
  \mathrm{Ry}^*(x,y) &=& \frac{q^2}{8 \pi \epsilon_0 \epsilon(x,y) a^*(x,y)}.
\end{eqnarray}
Material parameters have been taken from well established sources\cite{Vurgaftman} (the permittivity $\epsilon$ of GaAs and Al$_{0.3}$Ga$_{0.7}$As is taken as 13.18 and 12.24 respectively, the effective masses of $\Gamma$ valley electrons $m^*/m_0$ in GaAs and Al$_{0.3}$Ga$_{0.7}$As are assumed to be 0.067 and 0.092 respectively, while the conduction band offset between those two materials amounts to $\Delta E_c$=0.23 eV in our simulations).

We discretize both Eq. (\ref{SchroedingerEq}) and Eq. (\ref{PoissonEq}) in a finite--difference approach, allowing for different lattice spacings in orthogonal directions. 

Eq. (\ref{SchroedingerEq}) represents an eigenvalue problem, which we have to solve only in a restricted domain as the electron density equals zero for practical purposes far away from the interface. For computational efficiency reasons, the domain has been chosen as small as possible such that it does not change the numerical result compared to the solution within the whole heterostructure domain. Since only the lowest states are needed, the Lanczos algorithm\cite{Lanczos} is a suitable method for obtaining the eigenstates and their respective energies. We have chosen the freely available IETL\footnote{http://www.comp-phys.org/software/ietl/} implementation for this purpose. Having obtained a set of $k_{\rm max}$ eigenstates $\Psi_k$ allows us to calculate the electron density $n$ by integrating the Fermi distribution multiplied by the density of states of the orthogonal directions over energy. In the case of a two--dimensional simulation, it reads
\begin{equation}
  n=\sum_{k=1}^{k_{\rm max}} \left| \Psi_k \right|^2\,\int_{E_k}^\infty\left. \frac{\sqrt{2m^{*}}}{\pi \hbar \sqrt{E-E_k}}\frac{1}{1+\exp{\left(\frac{E-E_f}{k_b T}\right)}}\right. \,{\rm d}E,
\end{equation}
where $E_f$ and $E_k$ denote the Fermi energy and $k$--th eigenvalue respectively, while $T$ remains at liquid Helium temperatures of 4K throughout the investigation.

\begin{figure}
  \scalebox{0.3}{\includegraphics{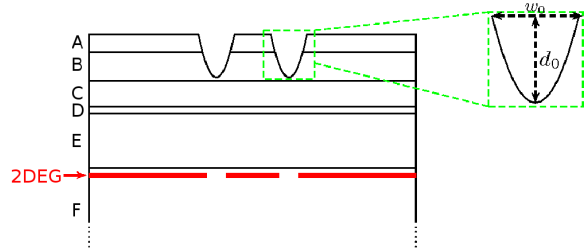}}
   \caption{(Color online) Schematic drawing of heterostructure layout: A: 5nm GaAs, B: 8nm Al$_{0.3}$Ga$_{0.7}$As, C: 7nm GaAs, D: 2nm Al$_{0.3}$Ga$_{0.7}$As n-doped, E: 15nm Al$_{0.3}$Ga$_{0.7}$As, F: bulk GaAs.}
    \label{LayoutFigure}
\end{figure}

For the solution of Eq. (\ref{PoissonEq}), we rely on the Biconjugate Gradient Stabilized Method.\cite{Vorst}
As boundary conditions for the system as depicted in Fig. \ref{LayoutFigure}, we demand a vanishing electric field in the bulk (bottom of Fig. 1) and Fermi level pinning on the semiconductor surface.\cite{Bardeen} The left and right side of Fig. 1 are connected by periodic boundary conditions, carefully paying attention to choose the lateral extension large enough to prevent images of the oxide line potential to have a considerable effect.

As both the Lanczos algorithm and the Biconjugate Gradient Stabilized Method mainly consist of matrix--vector products,
a parallel matrix--vector class has been implemented in order to achieve fast computation and ensure extensibility to three dimension which will impose significantly higher computational demands.

Starting with a trial potential, equations (\ref{PoissonEq}) and (\ref{SchroedingerEq}) are iteratively solved until a self--consistent solution is found.  To ensure convergence, suitable damping schemes\cite{Johnson} have to be applied in order to reach the desired equilibrium solution.

The heterostructure we want to investigate is schematically depicted in Fig. \ref{LayoutFigure}, following a design which has been realized and characterized experimentally.\cite{Held1,Held2,Held3} Layer D is n--doped in order to generate a 2DEG at the heterostructure interface. However, due to the fabrication process, it is known that the bulk GaAs (denoted F) is usually unintentionally p--doped. In our simulations, this results in an additional degree of freedom since for every background doping $N_p$ a modulation doping $N_n$ can be found that yields the experimentally measured sheet density $N_s=4.5\times 10^{11} \mathrm{cm}^{-2}$ of the 2DEG 37nm below the surface (see later discussion and Fig. \ref{DopingFig}). This relationship has been established using a one--dimensional self--consistent simulation (without oxide lines) along the growth direction.
We then apply the afore--mentioned numerical method to the structure as defined in Fig. \ref{LayoutFigure}, performing two--dimensional simulations in the plane perpendicular to the two parallel oxide lines in order to study the influence of parameter variations on the electronic width of the quantum wire. The latter forms in between the projections of the oxide lines on the 2DEG plane. In the experiment it has been established that the dimensions of the oxide line below and above the semiconductor surface are approximately the same. It was also demonstrated that the electronic properties of a device confined by oxide lines is not changed if the oxide lines are removed e.g. by HCl etching. Therefore we use in the following the term "oxide line" to describe the shape of the groove in the semiconductor surface which leads to a depletion of the electron gas below.

The precise shape and size of the oxide growth under the semiconductor surface cannot easily be probed experimentally. For the exact shape of the oxide line, we therefore choose a Gaussian--like form similar to what has been observed in the experiment.\cite{Fuhrer} Expressed in terms of the width $w_1$ as a function of depth $d$, it reads
\begin{equation} 
w_1(d)=2b\sqrt{-\ln{\left[1+\left(1-\frac{d}{d_0}\right)\left(\mathrm{e}^{-\left(\frac{w_0}{b}\right)^2}-1\right)\right]}},
 \label{OxLineShapeEq}
\end{equation}
where $d_0$ and $w_0$ represent the maximum depth and width, respectively, while the parameter $b$ characterizes the width at half depth. 

The main aim of this work is now to assess the influence of these unknown parameters on the electronic wire width quantitatively. In order to obtain independent parameters to vary we choose a definition of the lithographic width that does not depend on the exact oxide line shape. Following the experiment we therefore define the lithographic width $w_{li}$ as the horizontal distance of the oxide lines at a depth $d_0/2$ and investigate the influence of relative width $b/w_0$, maximum depth $d_0$ and the background doping $N_p$ respectively in the plane defined by the electronic and the lithographic width. The maximum width is kept constant at $w_0=200$nm throughout the simulations.

\section{Results}

Experimental electronic width values $w_{el}$\cite{Held2,Held3} as a function of lithographic width $w_{li}$ are plotted in Figures \ref{BckgrFig}--\ref{DepthFig} together with simulations for various parameter settings. In our simulation results, we define the electronic width $w_{el}$ as the spatial distance over which the self--consistent potential is below the Fermi energy. A linear interpolation of the experimental values yields a slope of $1.13 \pm 0.07$, while fittings for all simulations result in slopes in the range of $1.00 \pm 0.02$. The experimental error bars have been estimated based on available AFM scans and typical results for the wire width obtained from a fitting of the positions of the minima of the low-field magnetoresistance. The simulation results are within these error bounds.
\begin{figure}
 \scalebox{.34}{\rotatebox{-90}{\includegraphics{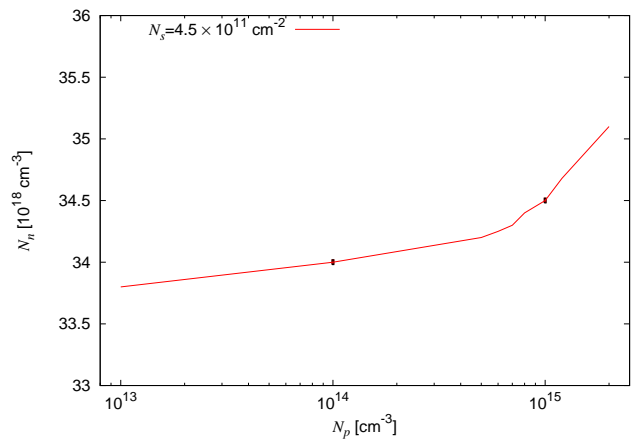}}}
 \caption{(Color online) Required modulation n--doping $N_n$ for a given background p--doping $N_p$ yielding a constant sheet density $N_s=4.5\times 10^{11}$ cm$^{-2}$ of the 2DEG.}
 \label{DopingFig}
\end{figure}

Since the background doping $N_p$, the maximum oxidation depth $d_0$ and the relative oxide line width $b/w_0$ are only known within certain ranges, we now investigate all three of them within their physically meaningful ranges and discuss their respective influences on the electronic width.

In order to assess the influence of the background doping uncertainty, two possible doping concentration combinations (corresponding to the two points marked in Fig. \ref{DopingFig}) are simulated and their effect on the electronic width is shown in Fig. \ref{BckgrFig}. The effect of the background doping turns out to be of minor importance.
\begin{figure}
 \scalebox{.34}{\rotatebox{-90}{\includegraphics{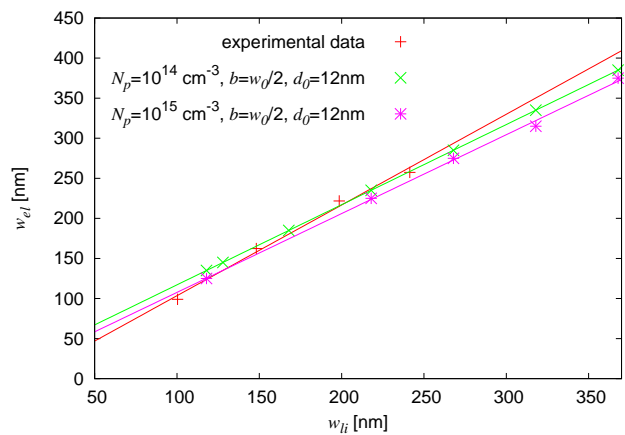}}}
 \caption{(Color online) Influence of background doping uncertainty on electronic width: Electronic width $w_{el}$ as a function of lithographic width $w_{li}$ for two simulated cases at constant $b=w_0/2$ and oxide line depth $d_0=12$nm for different background doping levels $N_p$ ($10^{14}$ cm$^{-3}$ and $10^{15}$ cm$^{-3}$) as well as experimental values. The symbols denote specifically calculated or measured values, the lines are linear fits to the corresponding data points.}
 \label{BckgrFig}
\end{figure}

In the following, we choose two extreme cases of possible oxide line shapes, $b=20w_0$ corresponding to a close to rectangular profile and $b=w_0/2$ which corresponds to the situation where the oxide depth decreases quickly from the center to the edge of the oxide line. All other parameters are kept constant. In Fig. \ref{ShapeFig}, we show the configuration at which the influence of $b$ is largest. It turns out that the choice of $b$ is more relevant for the comparison of experimental and simulated data than the background doping uncertainty.
\setlength{\unitlength}{0.01\linewidth}
\setlength{\fboxsep}{0pt}
\setlength{\fboxrule}{1.5pt}
\begin{figure}
 \begin{picture}(69,69)
  \put(-15,69){\scalebox{.34}{\rotatebox{-90}{\includegraphics{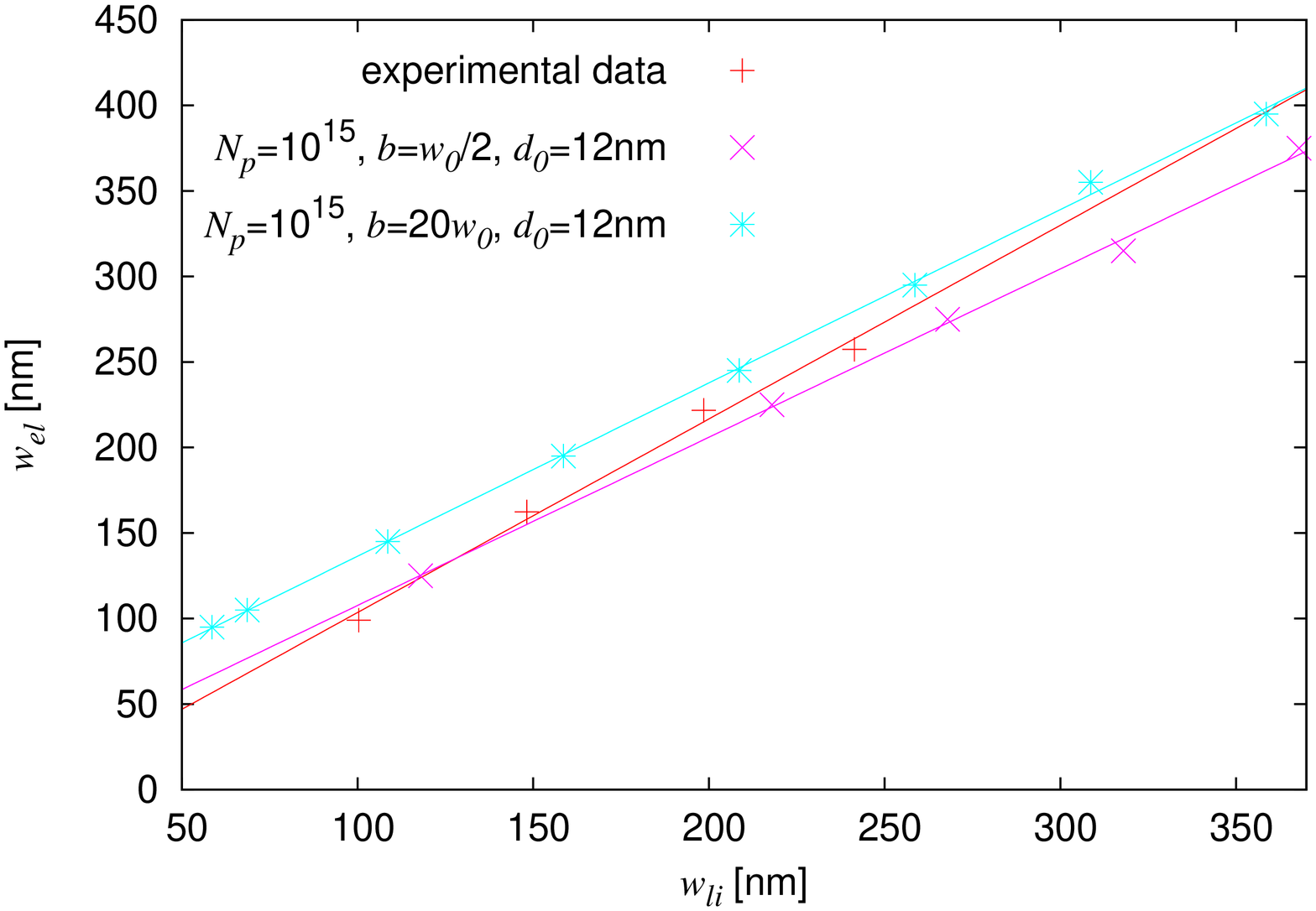}}}}
  \put(37.8,40.3){\scalebox{.15}{\rotatebox{-90}{\includegraphics{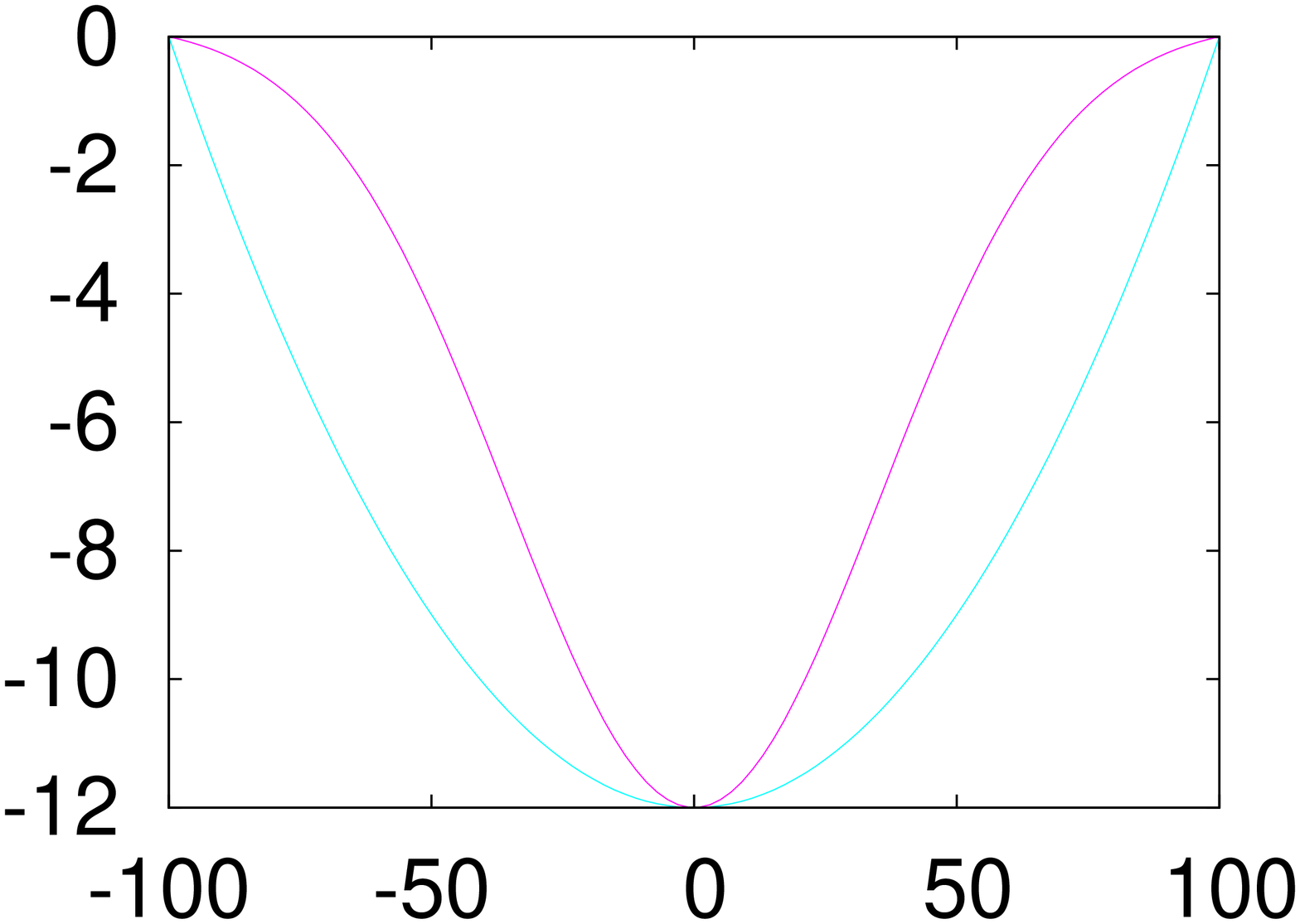}}}}
 \end{picture}
 \caption{(Color online) Influence of oxide line shape uncertainty on electronic width: Electronic width $w_{el}$ as a function of lithographic width $w_{li}$ for two simulated cases at constant background doping $N_p=10^{15}$ cm$^{-3}$ and oxide line depth $d_0=12$nm for different width parameters $b$ ($w_0/2$ and $20w_0$) as well as experimental values. The symbols denote specifically calculated or measured values, the lines are linear fits to the corresponding data points. Inset: geometrical shape of a single oxide line. The same color code as in the electronic width graph has been used to identify the different parameter sets (all axis units are nm).}
 \label{ShapeFig}
\end{figure}

Finally, all simulations have been carried out at two maximum oxide lines depths ($d_0=12$nm and $d_0=13$nm) which are most likely to represent the physical reality. Again, Fig. \ref{DepthFig} displays the line in configuration space at which the maximum depth exerts the most significant influence leaving all other parameters constant. This parameter turns out to be the crucial one affecting the electronic width as the 2DEG is depleted at larger lateral distances for larger values of $d_0$. Figure \ref{DepthFig} clearly shows that the minimum lithographic width for the population of the wire, i.e. the value of $w_{li}$ where $w_{el}$ goes to zero, increases with increasing oxide line depth $d_0$. For this special case a situation can be realized where the electronic width $w_{el}$ is larger than zero for a negative value of the lithographic width $w_{li}$. This peculiar configuration arises because of the definition of $w_{li}$. It means that the electron gas in the wire can laterally extend significantly under the oxidized areas, an effect which can be confirmed experimentally and is also known from split--gate defined quantum point contacts.
\begin{figure}
 \begin{picture}(69,69)
  \put(-15,69){\scalebox{.34}{\rotatebox{-90}{\includegraphics{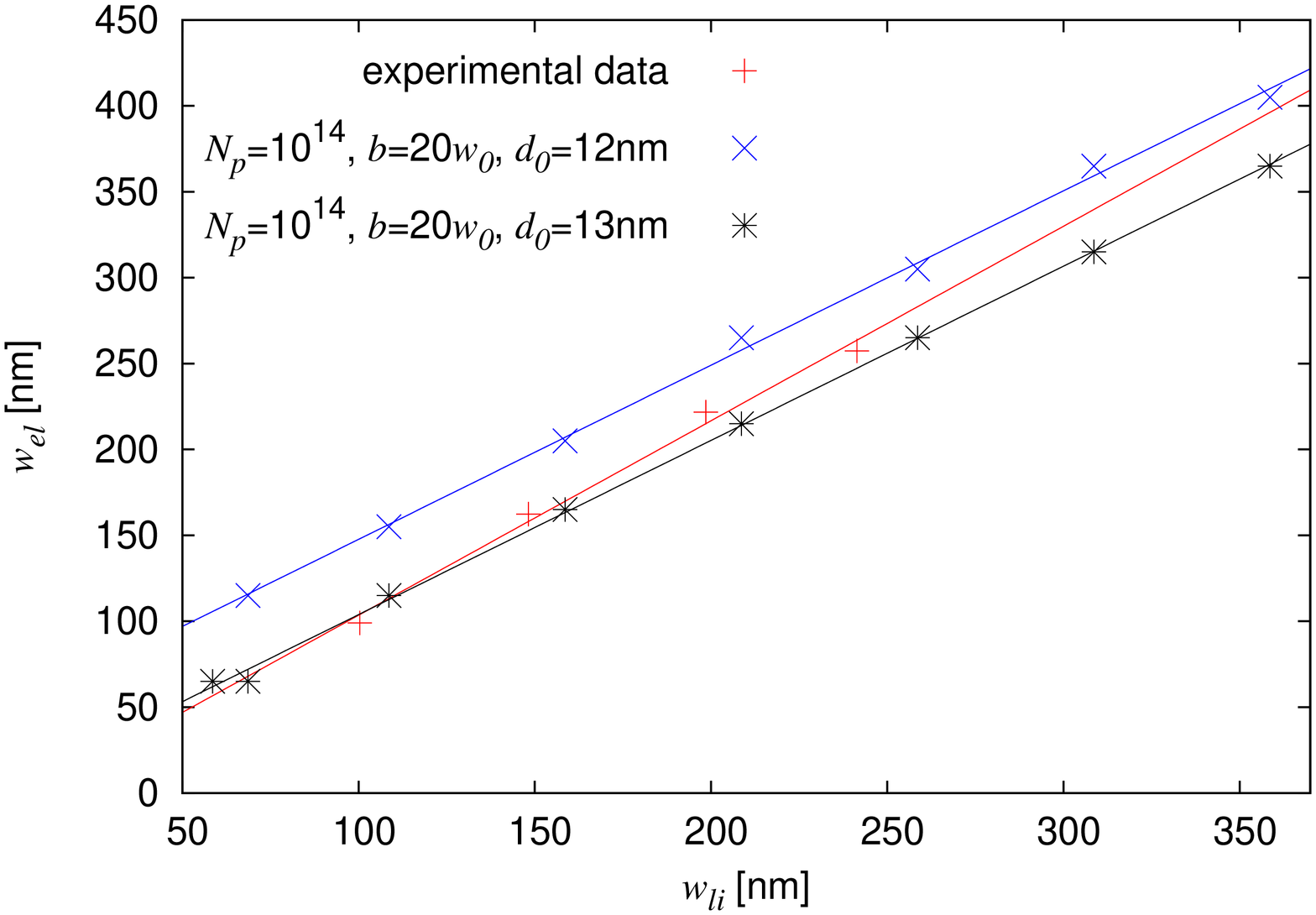}}}}
  \put(37.8,40.3){\scalebox{.15}{\rotatebox{-90}{\includegraphics{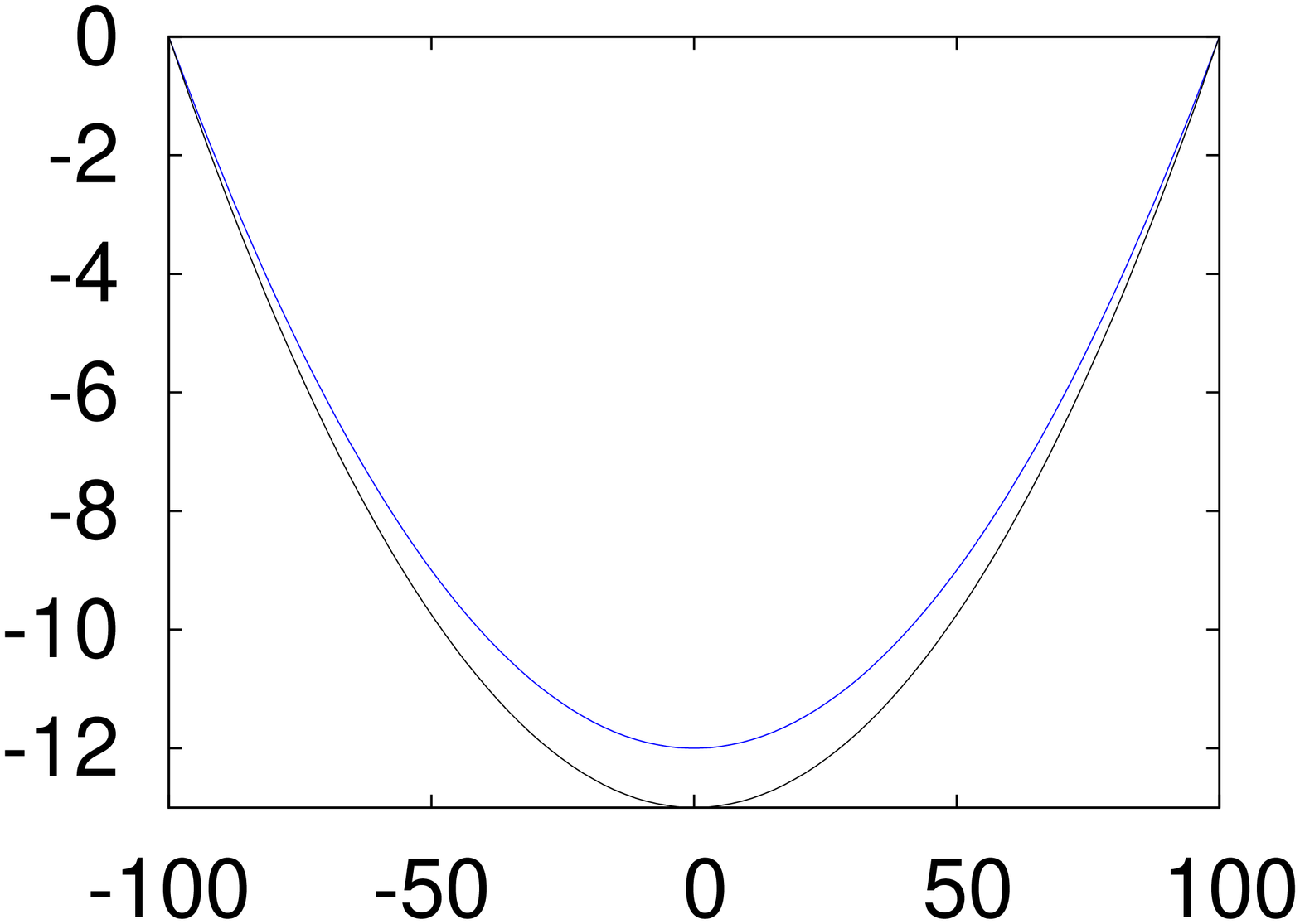}}}}
 \end{picture}
 \caption{(Color online) Influence of oxide line depth uncertainty on electronic width: Electronic width $w_{el}$ as a function of lithographic width $w_{li}$ for two simulated cases at constant background doping $N_p=10^{14}$ cm$^{-3}$ and width parameter $b=20w_0$ for different oxide line depths $d_0$ ($12$nm and $13$nm) as well as experimental values. The symbols denote specifically calculated or measured values, the lines are linear fits to the corresponding data points. Inset: geometrical shape of a single oxide line. The same color code as in the electronic width graph has been used to identify the different parameter sets (all axis units are nm).}
 \label{DepthFig}
\end{figure}

\begin{figure}
 \begin{picture}(69,69)
  \put(-15,69){\scalebox{.34}{\rotatebox{-90}{\includegraphics{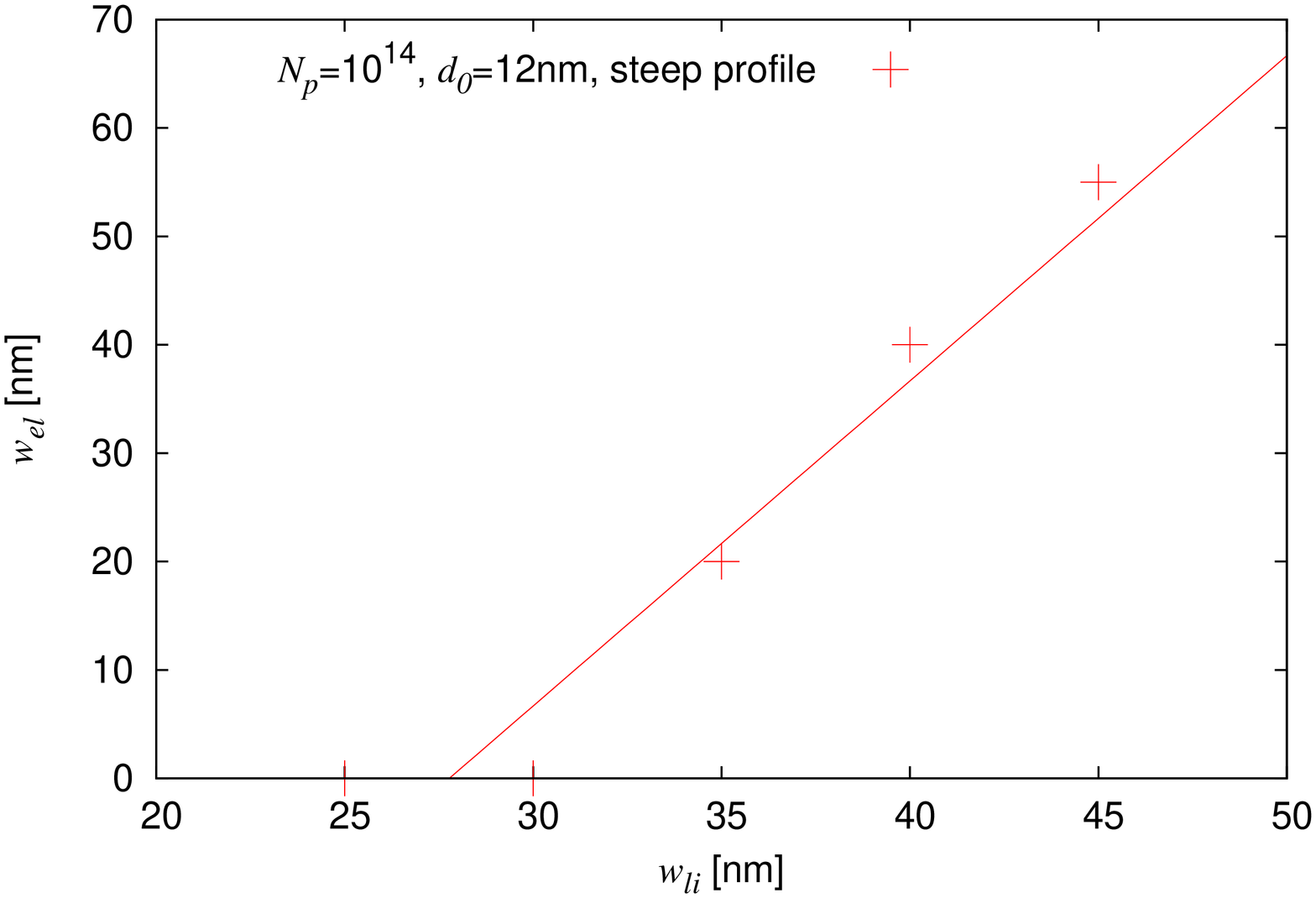}}}}
  \put(-3,62){\scalebox{.15}{\rotatebox{-90}{\includegraphics{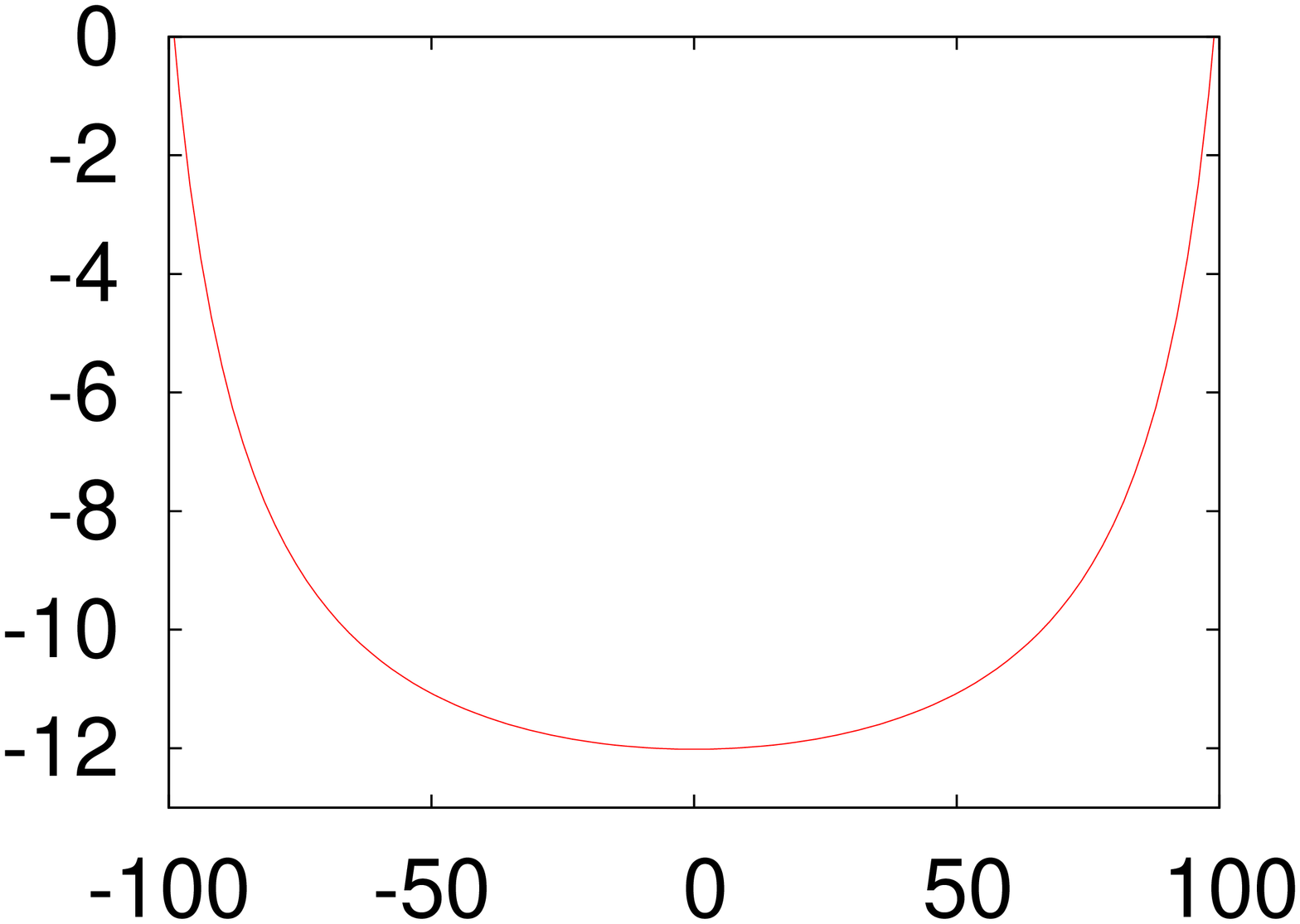}}}}
 \end{picture}
 \caption{(Color online) Electronic width $w_{el}$ vs. lithographic width $w_{li}$ curve in the small lithographic width limit at background doping $N_p=10^{14}$ cm$^{-3}$ and oxide line depth $d_0=12$ nm with a steep profile (see text). Inset: geometrical shape of a single oxide line (all axis units are nm).}
 \label{DownscaleFig}
\end{figure}

We note that our exploration of the parameter space given by experimental uncertainties does not affect the slope of the electronic width vs. lithographic width curve, but only adds constant offsets. This fact and the slope of unity we discovered agree with intuition in the case of a lithographic width exceeding all other physical length scales involved. However, on a smaller scale, there should be a higher slope due to screening as well as exchange and interaction effects, which cannot be observed using this oxide line shape because the lithographic width at half depth is still too wide for these phenomena to have a significant effect. In order to further investigate the behavior at smaller lithographic widths, we therefore implemented another shape allowing for extremely steep oxide line walls, which reads
\begin{equation}
  w_2(d)=2\sqrt{c_3^2-\frac{2 c_2 c_3}{d+c_1}}.
\end{equation}
We achieve steep walls with the desired oxide distance by choosing the constants $c_1$=$-1053/62$, $c_2$=$-55575\sqrt{62}/1922$, $c_3$=$450\sqrt{62}/31$. Fig. \ref{DownscaleFig} shows the resulting slope of $3 \pm 0.6$ for very close oxide lines. This extreme choice of parameters  is used to test  the theoretical limit rather than representing the actual geometry.

Extrapolation down to the intersection of the electronic width vs. lithographic width curve with the x--axis (lithographic width axis) allows one to read off the physically interesting depletion length. It turns out that moderate parameter variations can easily account for the difference between a positive depletion length (i.e. the quantum wire gets cut off at a finite oxide line distance) and a negative depletion length (it remains conducting).
\section{Conclusions}
Given that experimental error bars are not explicitly taken into account, the simulation data are in reasonable agreement with experiments. Of all parameters investigated, the oxidation depth turns out to be the most influential one. However, also the exact shape of the oxide line as well as the background doping uncertainty contribute to the overall error bounds (in decreasing order).

Given the exact physical setup of a nanostructure, the result of electronic width calculations is precisely determined by the well--known laws of quantum mechanics and electrodynamics. Since in experiments the exact shape is never exactly known and also differs from sample to sample, it is not straightforward to calculate the potential profile of more complex geometries.\cite{Fuhrer} We envision, however, that simulations using the methods presented in this paper will be useful to design novel structures and to obtain a better understanding on the electrostatic action of in-plane and top gates. Most importantly, simulations enable us to point out which parameters have the most significant effect. Better experimental control is therefore desirable for these parameters, in this particular case the oxide line depth.

\begin{acknowledgments}
We thank Jean-David Picon for discussions. All simulations were run on the Beowulf cluster \textit{Gonzales} operated by ETH Zurich and facilities at the Swiss National Supercomputing Center CSCS. This work was supported by ETH Research Grant TH-12 06-3 and the Studienstiftung des deutschen Volkes.
\end{acknowledgments}

\newpage 

\end{document}